\documentclass[%
 reprint,
nofootinbib,
 amsmath,amssymb,
 aps,
floatfix,
]{revtex4-2}

\usepackage{graphicx}
\usepackage{dcolumn}
\usepackage{bm}
\usepackage{xcolor}

\begin{document}

\title{Solving the Hubble tension at intermediate redshifts with dynamical dark energy}

\author{Isaac Tutusaus}
\email{isaac.tutusaus@irap.omp.eu}
\affiliation{%
Institut de Recherche en Astrophysique et Plan\'etologie (IRAP), Universit\'e de Toulouse, CNRS, UPS, CNES, 14 Av. Edouard Belin, 31400 Toulouse, France\\
}%

\author{Martin Kunz}
 \email{Martin.Kunz@unige.ch}
\affiliation{Département de Physique Théorique and Center for Astroparticle Physics,
Université de Genève, Quai E. Ansermet 24, CH-1211 Genève 4, Switzerland}

\author{Léo Favre}
\affiliation{Département de Physique Théorique and Center for Astroparticle Physics,
Université de Genève, Quai E. Ansermet 24, CH-1211 Genève 4, Switzerland}

\date{\today}

\begin{abstract}
The current expansion rate of the Universe, the Hubble constant $H_0$, is an important cosmological quantity. However, two different ways to measure its value do not agree -- building a low-redshift distance ladder leads to a higher value of $H_0$ than inferring it from high-redshift observations in a $\Lambda$CDM cosmology. Most approaches to solve this tension either act at very low redshift by modifying the local distance ladder, or at high redshift by introducing new physics that changes the normalization of the inverse distance ladder. Here we discuss a way to address the Hubble tension at intermediate redshifts instead. By keeping the low- and high-redshift normalizations unchanged, we find a violation of the distance duality in the redshift range where luminosity and angular diameter distances overlap. We `solve' this problem by introducing a redshift-dependent systematic effect that brings the luminosity distance into agreement with the angular diameter distance. The resulting expansion history is no longer compatible with $\Lambda$CDM, but this can be fixed with a dynamical dark energy component. In this way, we are able to solve the Hubble tension at intermediate redshifts.
\end{abstract}

\maketitle


\section{\label{Introduction}Introduction}

Since the discovery of the accelerated expansion 25 years ago \cite{SupernovaSearchTeam:1998fmf, SupernovaCosmologyProject:1998vns}, the cosmological standard model has been the Lambda-Cold-Dark-Matter ($\Lambda$CDM) model. $\Lambda$CDM has weathered many precision tests like the measurements of the anisotropies in the cosmic microwave background (CMB) by the Planck satellite \cite{Planck:2018nkj} or measurements of the distance-redshift relation using type-Ia supernovae (SN-Ia, e.g.\ \cite{Brout:2022vxf}) and baryonic acoustic oscillations (BAO, e.g.\ \cite{eBOSS:2020yzd}). 

With advancements in experimental sensitivity, it is reasonable to expect tensions between different data sets to emerge,   either due to unknown systematic effects in the data, or because we have reached a level of precision where we are able to uncover new physics beyond the standard model. Because of the latter, it is important to investigate and understand the source of such disagreements. The currently most significant tension between data sets comes from the determination of the Hubble constant $H_0$ that describes the late-time expansion rate of the Universe, see e.g.\ \cite{DiValentino:2021izs}: Local measurements with Cepheids and SN-Ia find $H_0 = 73.04 \pm 1.04$ km/s/Mpc~\cite{Riess:2021jrx} while the Planck data in combination with the $\Lambda$CDM model predicts $H_0= 67.7 \pm 0.42$ km/s/Mpc~\cite{Planck:2018vyg}, a number that is in agreement with the value obtained from BAO data combined with constraints on the baryon density $\omega_b$ from big-bang nucleosynthesis, without CMB anisotropies \cite{eBOSS:2020yzd}. The difference of more than five standard deviations between these measurements has led to a great interest in the community, as this might be the first sign of a breakdown of the $\Lambda$CDM model. Unfortunately, so far no systematic effects have been identified that can explain the tension fully, but also no model has been proposed that provides a convincing explanation for the observed discrepancy.

In this paper we focus particularly on intermediate redshifts where we have both luminosity distances from SN-Ia and angular diameter distances from BAO. As we will show below, these measurements are not compatible at equal redshifts. This is a puzzle, since they should simply be related by the distance duality relation (DDR) in any metric theory of gravity if photons are conserved \cite{Etherington1933,Bassett:2003vu}. Since SN-Ia are anchored at low redshifts and BAO at high redshifts, most proposals aim to change one of these anchors, either through systematic effects affecting the SN-Ia normalisation or through extensions of the standard model, like early dark energy, that affect the sound horizon at high $z$. Here we will instead assume that a systematic effect at intermediate redshifts changes the luminosity distance so that it agrees with the angular diameter distance. This is motivated by earlier studies that have considered a possible redshift dependence of the intrinsic SN-Ia luminosity. One example is the luminosity dependence on the local star formation rate. Although some analyses do not detect a significant dependence of the luminosity on this quantity, see e.g. \cite{Jones:2018vbn}, others do find a significant local environmental dependence of the intrinsic luminosity, see e.g. \cite{Kim:2023xed} and references therein. Given the redshift dependence of the star formation rate, this effect would lead to a redshift-dependent intrinisic SN-Ia luminosity. Another example is the dependence on the metallicity of the host galaxy. Metallicity evolves as a function of redshift, therefore, any luminosity dependence on the metallicity would introduce a redshift dependence, see e.g. \cite{Moreno-Raya:2016rlw,Moreno-Raya:2015jqq}. Observationally, a detection of a redshift-dependence of the intrinsic SN-Ia luminosity was recently claimed in~\cite{Perivolaropoulos:2023iqj}.
Allowing for a redshift-dependent intrinsic SN-Ia luminosity can have a large impact on the cosmological conclusions derived from the observations, see e.g. \cite{Tutusaus:2017ibk,Tutusaus:2018ulu}. As we will see, the ``deformation'' of the luminosity distance required to match the BAO distances then requires additionally late-time dark energy. The combination of a redshift-dependent systematic effect affecting the SN-Ia and a time-evolving dark energy was previously studied in \cite{Martinelli:2019krf}, but the authors found that the tension could not be fully removed. Very recently \cite{Gomez-Valent:2023uof} also considered scenarios including both systematic effects and an evolving dark energy component and found different possibilities to alleviate the tension. In this publication we argue that bringing the BAO and SN-Ia distances into agreement is not optional, and we design a scenario that uses a redshift-dependent intrinsic SN-Ia luminosity to enforce the DDR, together with a late-time dark energy to explain the resulting distance-redshift relation. In this way, our scenario solves the tension by construction.

\section{Cosmology and Data sets}

In this article, we will use distance data to constrain the expansion rate of the Universe. We will only consider flat models and choose the radiation density parameter $\Omega_r$ in agreement with the CMB temperature. 
The expansion rate as a function of redshift $z$ is then
\begin{equation}
H(z)^2 = H_0^2 \left(\Omega_r (1+z)^4+\Omega_m (1+z)^3 + \Omega_{\rm DE}(z)\right) \, .
\end{equation}
For the $\Lambda$CDM standard model, the dark energy density parameter is given by
\begin{equation}
\Omega_{\rm DE}(z) = (1-\Omega_r-\Omega_m) \, ,
\end{equation}
while for a general dark energy component with time-evolving equation of state parameter $w(z) = \bar{p}(z)/\bar{\rho}(z)$ it becomes
\begin{equation}
\Omega_{\rm DE}(z) = (1-\Omega_r-\Omega_m) \exp\left\{ 3 \int_0^z \frac{1+w(s)}{1+s} ds \right\} \, .
\end{equation}

\subsection{Type Ia supernovae}

Type Ia supernovae are standardizable candles, i.e.\ they measure the luminosity distance
\begin{equation}
D_L(z) = (1+z) \int_0^z \frac{ds}{H(s)} \,,
\end{equation}
which is usually expressed in terms of the distance modulus
\begin{equation}
\mu(z) = 5 \log_{10} \left(\frac{D_L(z)}{10 \mathrm{pc}}\right) \, .
\end{equation}
Here we use the Pantheon+ data \cite{Brout:2022vxf}, for which the observed distance modulus can be obtained as
\begin{equation}
    \mu_{\rm obs} = m_{B} - M \,,
\end{equation}
where $m_B$ is the observed magnitude from a light-curve fit and $M$ stands for the absolute luminosity of SN-Ia. The authors in \cite{Brout:2022vxf} provide the observed magnitudes after accounting for several systematic effects and marginalizing over different nuisance parameters, like the stretch of the light-curve, its colour, or the host galaxy mass.

We further include the SH0ES Cepheid host distance anchors \cite{Riess:2021jrx} to break the degeneracy between $M$ and $H_0$. The distance modulus residuals that enter the SN-Ia likelihood are then given by
\begin{align}
          \mu_{\rm obs}^i - \mu_{\rm Cepheid}^i\,,&\text{ if }i\in \text{Cepheid host,}\\
    \mu_{\rm obs}^i - \mu(z^i) \,,&\text{ otherwise,}
\end{align}

where $\mu_{\rm Cepheid}^i$ corresponds to the Cepheid calibrated host-galaxy distance.

When systematic effects introducing a redshift-dependent absolute magnitude of SN-Ia are included, we add an additional term into the observed distance modulus, leading to the residuals
\begin{align}
          \mu_{\rm obs}^i +\Delta \mu(z^i) - \mu_{\rm Cepheid}^i\,,&\text{ if }i\in \text{Cepheid host,}\\
    \mu_{\rm obs}^i +\Delta \mu(z^i)- \mu(z^i) \,,&\text{ otherwise.}
\end{align}

\subsection{Baryon Acoustic Oscillations}

Transverse baryon acoustic oscillations measure the comoving angular diameter distance,
\begin{equation}
    D_M(z) = D_L(z)/(1+z) \, .
\end{equation}
This relation is also called the distance duality and holds in any metric theory of gravity (see e.g.\ \cite{Bassett:2003vu}).

Like the SN-Ia distances, the BAO distances are also relative to a normalization, in this case the sound horizon scale at the drag epoch. This normalization can be affected by various mechanisms that act at high redshift, like early dark energy \cite{Poulin:2018cxd}. Here we assume that the physics at high redshift is not modified, so that we can use the standard values for that normalization.

In practice, we consider the legacy results of eBOSS \cite{eBOSS:2020yzd}, which contain the transverse BAO measurements from BOSS galaxies \cite{BOSS:2016wmc}, eBOSS LRGs \cite{Bautista:2020ahg,Gil-Marin:2020bct}, eBOSS ELGs \cite{Tamone:2020qrl,deMattia:2020fkb}, eBOSS quasars \cite{Hou:2020rse,Neveux:2020voa}, and the BOSS+eBOSS Lyman-$\alpha$ auto-correlation and cross-correlation \cite{duMasdesBourboux:2020pck}. In addition to the transverse measurements, we also include the radial BAO measurements 
from these data sets, as well as the angle-averaged measurement from the main galaxy sample of SDSS \cite{Ross:2014qpa,Howlett:2014opa}. We follow \cite{eBOSS:2020yzd} and use the covariance matrix for the BOSS galaxies measurements, as well as the covariance for the eBOSS LRGs and eBOSS quasars measurements. For the eBOSS ELGs and the Lyman-$\alpha$ forest measurements we employ instead the full likelihood provided by the collaboration, given its non-Gaussianity. The different data sets are considered to be independent from one another.

\subsection{Cosmic Microwave Background}

The temperature fluctuations observed with CMB anisotropies are not a simple distance measurement, and predicting them requires the use of a Boltzmann code. However, most of the information contained in the CMB can be compressed into a few numbers \cite{Mukherjee:2008kd}:
\begin{itemize}
    \item the CMB shift parameter $R = \sqrt{\Omega_m H_0^2} D_M(z_*)$ where $z_*$ is the redshift for which the optical depth becomes unity;
    \item the angular scale of the sound horizon at last scattering (equivalent to $\theta_*$), $\ell_a = \pi D_M(z_*)/r_s(z_*)$ where $r_s$ is the coming size of the sound horizon;
    \item as well as the baryon density $\omega_b$ and the scalar spectral index $n_s$. In this work, we keep $
    \omega_b$ but marginalize over $n_s$ since our other probes do not depend on it.
\end{itemize}
These numbers are effectively observables for the CMB. Their distribution is Gaussian to a good approximation, and is also valid for smooth dark energy components (or non-zero curvature) even when derived in a $\Lambda$CDM model. For the 2015 Planck data release, the mean values and covariance matrix was computed in \cite{Planck:2015bue}. Here, we derive an updated likelihood based on the Planck 2018 $\Lambda$CDM chains \cite{Planck:2018vyg} -- as shown in \cite{Mukherjee:2008kd}, the compressed likelihood does not depend on the model used. 

More specifically, we consider the 2018 plik TTTEEE+lowl+lowE Planck likelihood and assume a flat $\Lambda$CDM model. We further compute the radiation density parameter considering the approximation presented in \cite{Komatsu} with one massive neutrino of mass 0.06\,eV and a number of effective relativistic species in the early universe of $N_{\rm eff}=3.046$. We additionally consider the fitting function presented in \cite{Eisenstein:1997ik} to compute the ratio of baryon and photon densities that appears in the sound speed, assuming a CMB temperature of 2.728\,K. Finally, we set the redshift of the last scattering epoch at $z_*=1089$. With these approximations, we obtain the following mean values for the reduced parameters:\footnote{During the final stages of this work, an alternative compression of the CMB likelihood appeared in Appendix F of \cite{Rubin:2023ovl}, based on flat $w$CDM chains from Planck.}
\begin{equation}
    (\ell_a,R,\omega_b) = (301.205,1.7478,0.02236)\,,
\end{equation}
together with the covariance
\begin{equation}
    \text{cov}=10^{-5}
\times\left(
    \begin{array}{ccc}
         712.9489 & -1.7156 &  0.2656\\
         -1.7156 &  2.1569 & -0.0459\\
         0.2656 & -0.0459 &  0.0023\end{array}
         \right)\,.
\end{equation}

\subsection{Compatibility of the distances}

\begin{figure}
\includegraphics[width=0.99\linewidth]{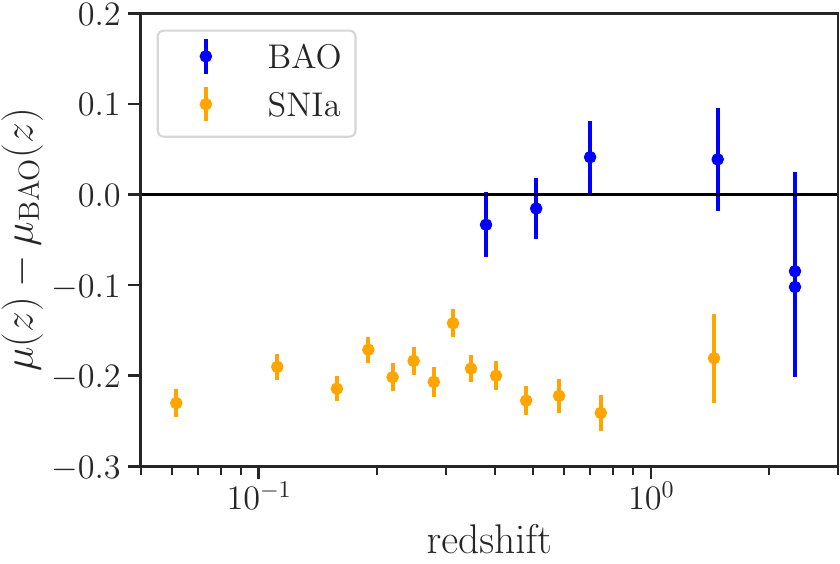}
\caption{Residuals of the distance moduli for SN-Ia and transverse BAO with respect to the BAO best-fit $\Lambda$CDM model. The discrepancy between the two data sets illustrates the Hubble tension.}
\label{fig:distances}
\end{figure}

In Fig.\ \ref{fig:distances} we show the redshift-binned luminosity distance moduli of SN-Ia, together with the BAO distance moduli. The latter have been obtained from the (comoving) BAO distances multiplied by the factor $(1+z)$ that should transform them to luminosity distances according to the distance duality. It is obvious that the distances do not agree in the shared redshift range. We note that such an inconsistency has already been identified in previous analyses \cite{Raveri:2023zmr,Pogosian:2021mcs,Raveri:2021dbu}, where the authors reconstructed an energy component to better fit both data sets at the same time. However, while dark energy can improve the overall fit, it {\em cannot} solve the discrepancy between the distance measurements, as it affects all distances. Similar results were found in \cite{Keeley:2022ojz,Linder:2023ndd}, where, even with a very flexible parameterization for the dark energy equation of state, no solution to the Hubble tension could be found with new physics at low-redshift. For this reason, we instead restore the distance duality by invoking the presence of systematic uncertainties to reconcile the two data sets without modifying their anchors. We also note that the use of the cosmic distance duality has been considered recently to constrain a possible redshift evolution of the intrinsic SN-Ia luminosity \cite{Hu:2023kgk}. The authors of this work combined SN-Ia and strong lensing measurements, but the distance measurements of these two data sets are compatible and led to no detection of a redshift evolution of the intrinsic SN-Ia luminosity.

\section{Schematic analysis}

\begin{figure}
\includegraphics[width=0.99\linewidth]{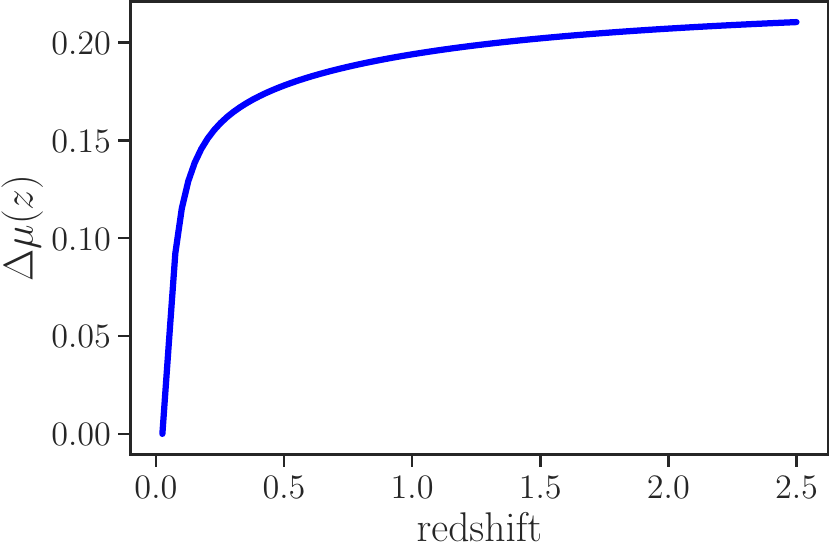}
\caption{The change of SN-Ia distance moduli, $\Delta\mu(z)$, necessary to reconcile BAO and SN-Ia distances.} 
\label{fig:dmu}
\end{figure}

Here we concoct a simple, schematic example of how to address the $H_0$ discrepancy at intermediate redshifts.
Although the two distances do not agree, each distance agrees separately with a $\Lambda$CDM model, for different values of $H_0$. We assume that there is a systematic effect that affects the observed luminosity distances so that they transition from distances corresponding to one value of $H_0$ to another. More precisely, we use a transition function $s(z)$ that changes from $s=0$ to $s=1$,
\begin{equation}
s(z) = \frac{1}{1 + e^{- (z -d) c}} \label{eq:dmu}\,,
\end{equation}
where $c,d$ are parameters that describe the transition: $d$ the location in $z$ and $c$ the rapidity. We then assume that there are two $\Lambda$CDM expansion rates given at late times by
\begin{equation}
    H^{[i]}(z) = H^{[i]}_0 \left(\Omega_m^{[i]} (1+z)^3 + (1-\Omega_m^{[i]}) \right)^{1/2} \label{eq15}
\end{equation}
for two different values of $H^{[i]}_0,\Omega_m^{[i]}, \, i = 1,2$.
The effective expansion rate for the luminosity distance is then
\begin{equation}
    H_{\rm SN-Ia}(z) =  (1-s(z)) H^{[1]}(z) + s(z) H^{[2]}(z) \, . \label{eq:hz}
\end{equation}
We compute the luminosity distance and the distance modulus from this expansion rate as well as from $H^{[1]}(z)$ -- which corresponds to SN-Ia without systematic effects, and consider the difference of the distance moduli $\Delta \mu(z)$ as the necessary systematic effect that can reconcile the data sets. We show an example for $\Delta\mu(z)$ that is able to fit the data well in Fig.\ \ref{fig:dmu}.

\begin{figure}
\includegraphics[width=0.99\linewidth]{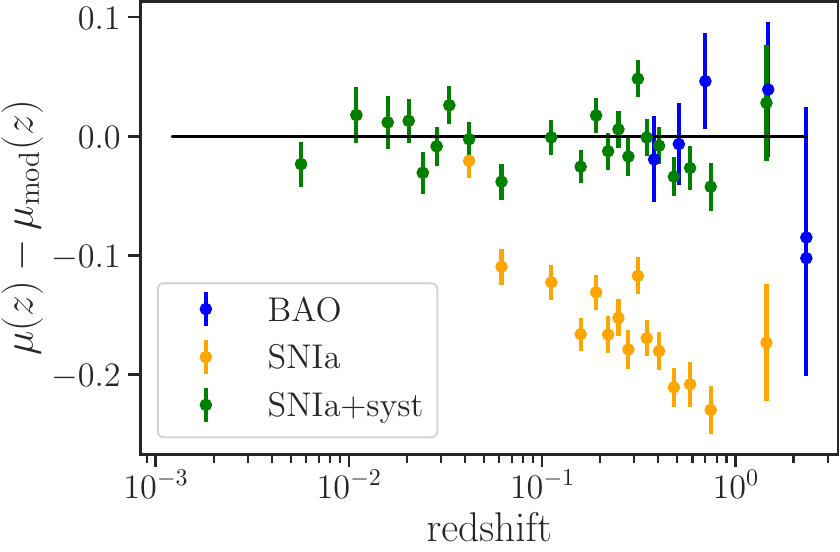}
\caption{A comparison of distance moduli for SN-Ia and BAO distances (blue points), relative to the best fit $w$CDM model for the (SN-Ia+syst)+BAO+CMB points (black line).  
The orange points show the SN-Ia distance moduli from the Pantheon+ data without systematic effect.
The green points show the result of applying the correction shown in Fig.\ \ref{fig:dmu}. Clearly the green points are in much better agreement with the blue BAO points than the orange points. This figure uses a dark energy component with the $w(z)$ shown in Fig.\ \ref{fig:wz}.}
\label{fig:mumod}
\end{figure}

The resulting luminosity distance is able to reconcile SN-Ia and BAO distances, see Fig.\ \ref{fig:mumod}. However, it does not correspond to the expansion rate of a $\Lambda$CDM model, because of the transition. But for any expansion rate $H(z)$ and a given $\Omega_m$, there is a dark energy equation of state $w(z)$ that reproduces $H(z)$ \cite{Kunz:2007rk}. Since we have an explicit form for $H(z)$ from Eq.\ \eqref{eq:hz}, we can easily derive the corresponding equation of state parameter. By adjusting $\Omega_m$, we find that we can bring it into a form where $w(z)$ is constant at small and large redshifts, and has a bump around the transition redshift $z=d$. This suggests that we should be able to model such a shape with two asymptotic values, $w_1$ and $w_2$, and a Gaussian peak,
\begin{equation}
w(z) = w_1 (1-s(z)) + w_2 s(z) + 
    u e^{-\frac{1}{2} (z - d)^2/v^2} \, .
\end{equation}
Here $u$ is the amplitude of the bump and $v$ controls the width, while the location is given by $d$.

\begin{figure}
\includegraphics[width=0.99\linewidth]{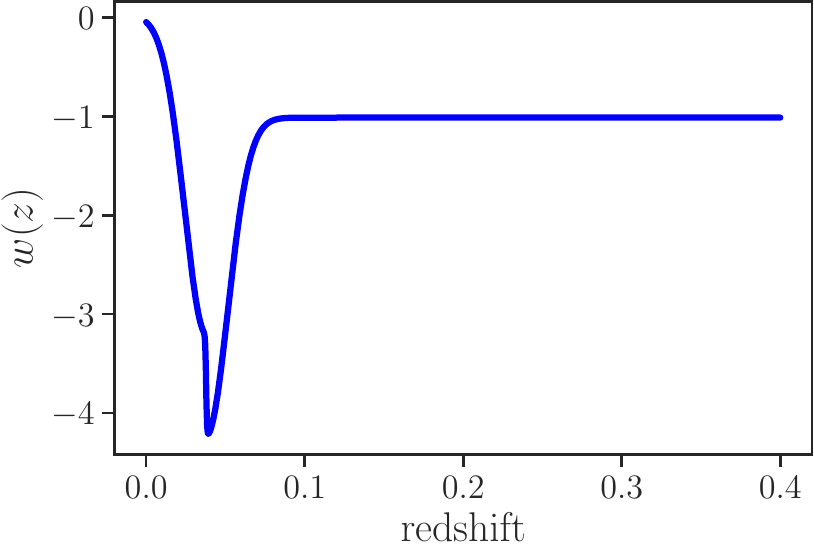}
\caption{The equation of state parameter $w(z)$ required to fit the full Hubble diagram that corresponds to the green and blue points in Fig.\ \ref{fig:mumod}.} 
\label{fig:wz}
\end{figure}

Summarizing, we consider a cosmological model that accounts for some systematic effects through $\Delta \mu(z)$ and dark energy characterized by an equation of state $w(z)$. The former can be extracted from the difference of distance moduli from Eqs.\,(\ref{eq15},\ref{eq:hz}). Therefore, the systematic effects are parametrized by the nuisance parameters $c,d$, as well as the Hubble constant from the SN-Ia best-fit, $H_0^{[1]}$, the Hubble constant from the BAO+CMB best-fit, $H_0^{[2]}\equiv H_0^{[1]}+\Delta H_0$, the matter density from the SN-Ia best-fit, $\Omega_m^{[1]}$, and the matter density from the BAO+CMB best-fit, $\Omega_m^{[2]}$. Concerning $w(z)$, we add the two amplitudes, $w_1,w_2$, and the nuisance parameters $u,v$. Finally, we also need to account for the Hubble constant of our model, $H_0$, as well as the matter density, $\Omega_m$, the baryon density, $\Omega_b$, and the absolute magnitude of SN-Ia, $M$. This leads to a total number of 14 free parameters. However, in order to simplify the analysis and avoid large degeneracies, we fix $H_0^{[1]}=73.4$\,km/s/Mpc, $\Omega_m^{[1]}=0.331$, $\Delta H_0=5.8$\,km/s/Mpc, and $\Omega_m^{[2]}=0.312$, which correspond to the best-fit values obtained from the fit of a $\Lambda$CDM model to the individual data sets. Maximizing the goodness of fit (leading to $\Omega_m \approx 0.27$), we obtain the shape shown in Fig.\ \ref{fig:wz}, where $w$ remains close to $-1$ down to $z\approx 0.1$ and then exhibits a `phantom bump' around $z \approx 0.05$ (in agreement with the general considerations of \cite{Heisenberg:2022lob}). 

Overall our scenario improves the goodness of fit of the data sets by $\Delta\chi^2 \approx 40$ and brings SN-Ia, BAO and CMB data into agreement, at the price of introducing two new ingredients at low redshifts, a systematic effect affecting the SN-Ia distances and a late-time dark energy. While requiring two ingredients may not appear compelling, each one of them is not very exotic per se, and a key point of this analysis is to show that the combination of the two provides an alternative scenario that provides a good fit to the data.

\section{\label{Conclusion}Conclusion}

In this paper we show that it is possible in principle to fit simultaneously the normalized supernova data, the BAO data and the main features of the CMB with late-time dark energy, if there is a systematic effect affecting the supernova luminosities, even if that effect does not change the late-time $H_0$. If we neither want to change the normalisation of the SN-Ia, nor introduce a deviation from $\Lambda$CDM at high redshifts, then the presence of such a systematic effect is effectively unavoidable as the distance duality requires luminosity and angular diameter distances to agree. 

This systematic effect needs to reach about 0.2 magnitudes at high redshift, and needs to reach about 80\% of this change by redshift $z\approx 0.3$ in order to bring the SN-Ia distances into agreement with the BAO distances.
One future way to test for such a systematic effect could be a comparison between the gravitational wave (GW) inferred luminosity distance and the SN-Ia luminosity distance as soon as we are able to observe a GW event in a galaxy that also hosted a SN-Ia \cite{Gupta:2019okl}. This would be a crucial test whether the $H_0$ tension can be explained by a mechanism that affects neither the distance anchors nor the primordial universe.

The systematic effect changes the true distance-redshift relation at low $z$ away from $\Lambda$CDM, requiring now the existence of a late-time evolving dark energy component. The best-fit $w(z)$ from our schematic analysis shows a relatively large excursion to $w \approx -4$ below $z \approx 0.1$. This might be visible in other data sets, but it is challenging to observe as it happens at very low redshifts.

Another potential tension concerns the amplitude of fluctuations on scales of $8 h^{-1}$ Mpc, $\sigma_8$. The value of $\sigma_8$ inferred from the CMB in a $\Lambda$CDM model is higher than the value found by weak lensing measurements (see e.g. \cite{KiDS:2020suj,DES:2021vln,DES:2021bvc}), 
and some mechanisms that aim to solve the Hubble tension aggravate the $\sigma_8$ tension \cite{Schoneberg:2021qvd}. 
We computed the value of $\sigma_8$ in our scenario where $w$ varies, and it turns out to be about 5\% higher than in the corresponding $\Lambda$CDM case ($\sigma_8\approx 0.834$ compared to $\sigma_8 \approx 0.793$). However, our preferred matter density is also somewhat lower than in the standard model, so that our value of $S_8\approx 0.79$ is compatible with the weak lensing observations of \cite{DES:2021wwk} at 1$\sigma$.

\begin{acknowledgments}
MK acknowledges financial support from the Swiss National Science Foundation.
\end{acknowledgments}

\bibliography{mybib}

\end{document}